\documentclass[pre, 11pt, superscriptaddress, amsmath, amssymb, floatfix]{revtex4}

\usepackage{graphicx}%
\usepackage{dcolumn} %
\usepackage{bm}      %
\usepackage{subfigure}
\usepackage{amsfonts}
\usepackage[usenames]{color}
\usepackage{rotating}
\usepackage{fontenc}
\usepackage{cancel}
\usepackage{amsthm}
\usepackage{hyperref}
\usepackage[normalem]{ulem}

\definecolor{darkred}{RGB}{139,0,0}
\definecolor{chartreuse}{RGB}{127,255,0}
\definecolor{goldenrod}{RGB}{218,165,32}
\definecolor{gray}{RGB}{127,127,127}

\definecolor{Magenta}{RGB}{255, 0,255}
\definecolor{Orange}{RGB}{255,165, 0}
\definecolor{Gray}{RGB}{127,127,127}

\newcommand{\be}{\begin{equation}}
\newcommand{\ee}{\end{equation}}
\newcommand{\bea}{\begin{eqnarray}}
\newcommand{\eea}{\end{eqnarray}}
\newcommand{\bw}{\begin{widetext}}
\newcommand{\ew}{\end{widetext}}
\newcommand{\mm}{\mathrm}

\newcommand{\bi}{\begin{itemize}}
\newcommand{\ei}{\end{itemize}}

\begin{document}
\title{Core percolation on complex networks}
\author{Yang-Yu Liu}  
\affiliation{Center for Complex Network Research and Department of
  Physics,Northeastern University, Boston, Massachusetts 02115, USA} 
\affiliation{Center for Cancer Systems Biology, Dana-Farber Cancer Institute, Boston, Massachusetts 02115, USA}
\author{Endre Cs\'oka}
\affiliation{E\"otv\"os Lor\'and University, Budapest, Hungary}

\author{Haijun Zhou}
\affiliation{State Key Laboratory of Theoretical Physics, Institute of Theoretical Physics, Chinese Academy of Sciences, Beijing 100190, China}

\author{M\'arton P\'osfai}
\affiliation{Center for Complex Network Research and Department of
  Physics,Northeastern University, Boston, Massachusetts 02115, USA} 
\affiliation{Department of Theoretical Physics, Budapest
   University of Technology and Economics, Budapest, Hungary}
\affiliation{Department of Physics of Complex Systems, E\"otv\"os
  Lor\'and University, Budapest, Hungary}
\date{\today}

\begin{abstract}
As a fundamental structural transition in complex
networks, core percolation is related to a wide range of important 
problems. %
Yet, previous theoretical studies %
of core percolation have been focusing on the classical Erd\H{o}s-R\'enyi
random networks with Poisson degree
distribution, %
which are quite unlike many real-world networks with scale-free or
fat-tailed degree distributions. %
Here we show that core percolation can be analytically studied for
complex networks with arbitrary degree distributions. 
We derive the condition for core percolation and find that purely
scale-free networks have no core for any degree exponents. 
We show that for undirected networks if core percolation occurs then
it is always continuous while %
for directed networks it becomes discontinuous when the in- and
out-degree distributions are different. 
We also apply our theory to real-world directed networks and find,
surprisingly, that they often have much larger core sizes as compared
to random models. 
These findings would help us better understand the interesting
interplay between the structural and dynamical properties of complex
networks. 
\end{abstract} 

\maketitle

Network science has emerged as a prominent field
in complex system research, which provides us a novel perspective
to better understand
complexity\cite{Strogatz-Nature-01,Barabasi-NaturePhysics-11,Vespignani-NP-11}.  
In the last decade considerable advances about structural and
dynamical properties of complex networks have been
made\cite{Albert-RMP-02,Newman-SIAM-03,Dorogovtsev-RMP-08}. 
Among them, structural transitions in networks were extensively 
studied due to their big impacts on numerous dynamical processes on
networks. 
Particularly interesting are the emergence of a giant connected
component\cite{Erdos-PMIHAS-60,Bollobas-Book-01,Callaway-PRL-00,Newman-PRE-01} , $k$-core
percolation\cite{Pittel-JCT-96,Dorogovtsev-PRL-06,Goltsev-PRE-06}, $k$-clique
percolation\cite{Palla-Nature-05,Derenyi-PRL-05}, and explosive percolation\cite{Achliopta-Science-09,Costa-PRL-10,Riordan-Science-11}. These structural transitions affect many
properties of networks, e.g. robustness and resilience to
breakdowns\cite{Cohen-PRL-00,Callaway-PRL-00,Albert-Nature-00},
cascading failure in interdependent
networks\cite{Buldyrev-Nature-10,Parshani-PRL-10,Gao-NP-11,Gao-PRL-11},
epidemic and information spreading on socio-technical
systems\cite{Pastor-Satorras-PRL-01,Kitsak-NP-10,Vespignani-NP-11}.  
Recent work on network controllability reveals another interesting
interplay between the structural and dynamical properties of complex
networks\cite{Liu-Nature-11,Nepusz-NP-12,Posfai-2012}. It was
found that the robustness of network controllability is closely
related to the presence of the \emph{core} in the network\cite{Liu-Nature-11,Jia-2012}.  
Actually, core percolation has also been related to many other
interesting problems, including conductor-insulator
transitions\cite{Bauer-PRL-01,Bauer-EPJB-01} and some classical combinatorial
optimization problems, e.g. maximum
matching\cite{Karp-IEEE-81,Zhou-arXiv-03,Zdeborova-JSM-06} and vertex
cover\cite{Weigt-PRL-00,Zhou-EPJB-03,Hartmann-JPA-03}. 

The core of a undirected network is defined as a spanned subgraph
which remains in the network after the following greedy leaf removal (GLR)
procedure\cite{Karp-IEEE-81,Bauer-EPJB-01}: As long as the network has
leaves, i.e. nodes of degree 1, choose an arbitrary leaf $v_1$, and its
neighbor $v_2$, and remove them together with all the edges incident
with $v_2$. Finally, we remove all isolated nodes. It can be proven
that the resulting graph is independent of the order of
removals\cite{Bauer-EPJB-01}. 
Note that the core described above is fundamentally different from the
$k$-core of a network. The latter is defined to be the maximal
subgraph having minimum node degree of at least $k$, which can be
obtained by iteratively removing nodes of degree less than
$k$.   
Apparently, the GLR procedure described above is more destructive than the
node removal procedure used to obtain the $2$-core (see Fig.\ref{fig:schematic}a). 
In studying the robustness of controllability for general directed
networks, the GLR procedure has been extended to calculate the core of
directed networks\cite{Liu-Nature-11}. 
We first transform a directed network $\mathcal{G}$ to its bipartite graph
representation $\mathcal{B}$ by splitting each node $v$ into two nodes
$v^+$ (upper) and $v^-$ (lower), and we connect $v_1^+$ to $v_2^-$ in
$\mathcal{B}$ if there is a link $(v_1 \to v_2)$ in $\mathcal{G}$. The
core of a directed network $\mathcal{G}$ can then be defined as the
core of its corresponding bipartite graph $\mathcal{B}$ obtained by
applying GLR to $\mathcal{B}$ as if $\mathcal{B}$ is a unipartite
undirected network.

One can easily tell whether the core exists in two very special cases: (1) If a
network has no cycles, i.e. a tree or a forest (a disjoint union of
trees), then eventually all nodes will be removed, hence no core. For
example, the Barab\'{a}si-Albert (BA) model with parameter $m=1$ yields a tree
network, hence no core exists.  (2) If a network has no leaf
nodes, e.g. regular graphs with all nodes having the same degree
$k>1$ or the networks generated by the BA model with
$m>1$, then the GLR procedure will not even be initiated, hence all the nodes belong to
the core. 

Except those two special cases, no general rules have been proposed
to predict the existence of the core for an arbitrarily complex
network. 
Previous theoretical studies focused on undirected
Erd\H{o}s-R\'enyi (ER) random graph. 
It has been show that for mean degree $c \le e = 2.7182818\ldots$, the
core is small (zero asymptotically), whereas for $c > e$ the core
covers a finite fraction of all the nodes\cite{Karp-IEEE-81,Bauer-EPJB-01,Zhou-Book-12}. 
In other words, core percolation occurs at the critical point $c*=e$. 
More interestingly, it has been suggested that in ER random graph core
percolation coincides with the changes of the solution-space structure 
of the vertex cover
problem\cite{Weigt-PRL-00,Hartmann-JPA-03,Hartmann-JPC-08}, which is 
one of the basic NP-complete optimization problems\cite{Garey-Book-79}. 
Also, for $c\le e$ the typical running time of an algorithm for finding the
minimum vertex cover is polynomial\cite{Weigt-PRL-00,Bauer-EPJB-01},
while for $c>e$, one needs typically
exponential running time\cite{Barthel-PRE-04}. 
Hence, core percolation also coincides with an ``easy-hard
transition'' of the typical computational
complexity\cite{Hartmann-JPA-03,Hartmann-JPC-08}. 

Despite the results on undirected ER random networks and the
importance of understanding the intriguing interplay between core
percolation and other problems, we lack a
systematic study and a general theory of core percolation for both
undirected and directed random networks with arbitrary degree
distributions.

\section{Analytical framework}
\begin{figure}
\begin{center}
\includegraphics[width=0.75\textwidth]{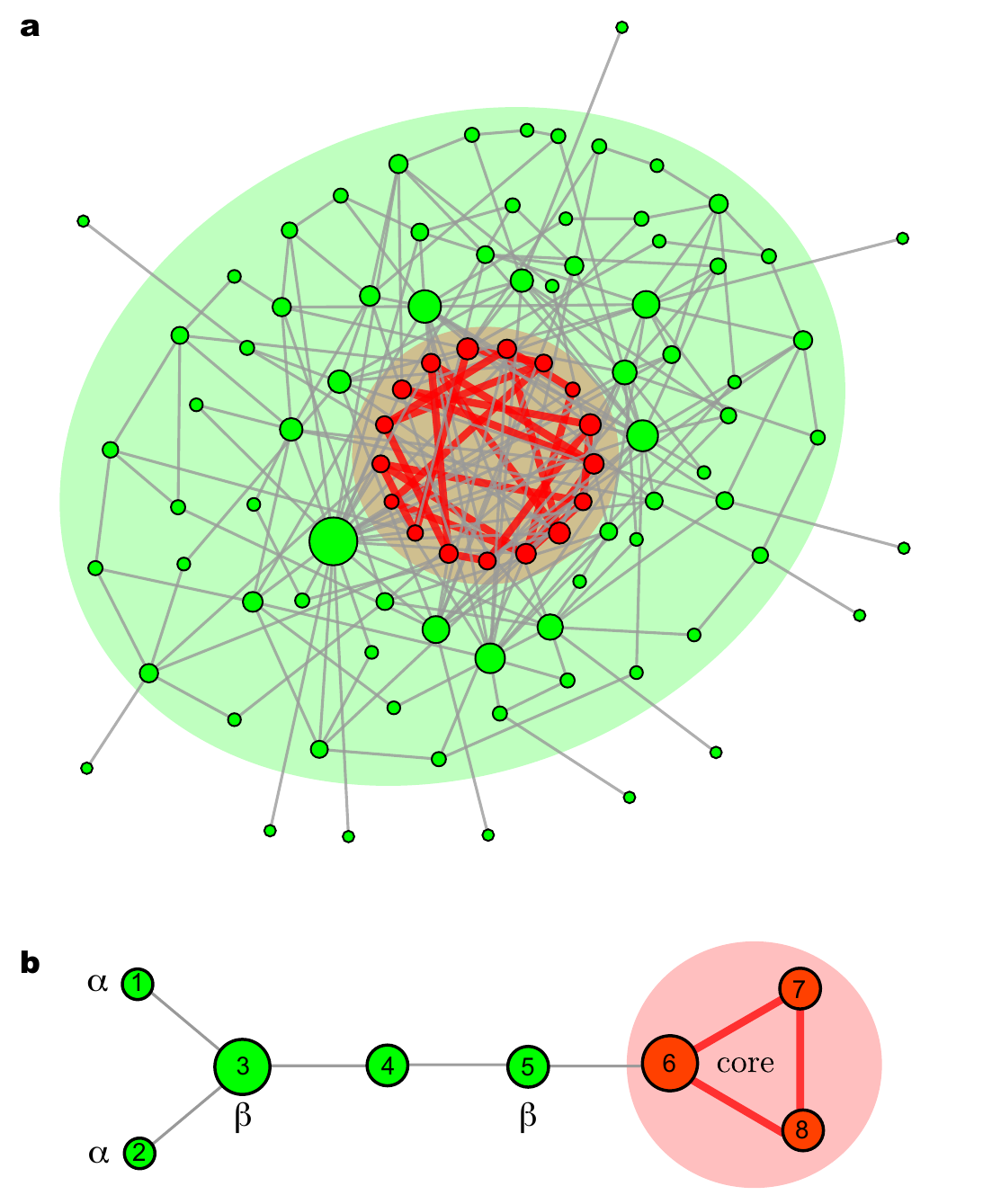}
\end{center}
\caption{{\bf The core of a small network.} {\bf a}, The core (highlighted in
  red) obtained after the greedy leaf removal procedure is fundamentally
  different from the $2$-core (highlighted in green) obtained by
  iteratively removing nodes of degree less than $2$. %
  The 2-core contains the core, whereas the opposite is not true. 
  Size of nodes are roughly proportional to the degree of nodes. 
  {\bf b}, Removal categories of nodes according to how they can be
  removed during the greedy leaf removal procedure. 
  Red nodes are non-removable, i.e. they belong to the core.
  Green nodes are removable: nodes $v_1$ and $v_2$ are
  $\alpha$-removable; nodes $v_3$ and $v_5$ are $\beta$-removable.   
  White node $v_4$ is removable but it is neither
  $\alpha$-removable nor $\beta$-removable. Node $v_5$ is
  $\beta$-removable because $v_4$ will become a leaf node after
  removing node $v_1$ (or $v_2$) together with $v_3$. 
}\label{fig:schematic}
\end{figure}

We propose the following analytical framework to study core
percolation on random networks with arbitrary degree distributions. 
We first categorize the nodes according to how they can be
removed during the GLR procedure. We define the following categories:  
(1) $\alpha$-removable: nodes that can become isolated (e.g. $v_1$ and
$v_2$ in Fig.\ref{fig:schematic}b); (2)
$\beta$-removable: nodes that can become a neighbor of a leaf
(e.g. $v_3$ and $v_5$ in Fig.\ref{fig:schematic}b); (3)
non-removable: nodes that cannot be removed and hence belong to the
core (e.g. $v_6, v_7$ and $v_8$ in Fig.\ref{fig:schematic}b). 
While the core is independent of the order the leaves are
removed\cite{Bauer-EPJB-01}, the specific way a node is removed may
depend on this order, but it can be proven that no node can be both
$\alpha$-removable and $\beta$-removable at the same time. 
Now we consider an uncorrelated random network with arbitrary 
degree distribution $P(k)$\cite{Newman-PRE-01,Molloy-RSA-95}. Assuming 
that in each removable category the removal of a random node can be made
locally, %
we can determine the category of a node $v$ in a network
$\mathcal{G}$ by the categories of its neighbors in $\mathcal{G}
\setminus v$, i.e.  the subgraph of $\mathcal{G}$ with node $v$ and
all its edges removed, using the following rules:  
(1) $\alpha$-removable: all neighbors are $\beta$-removable; 
(2) $\beta$-removable: at least one neighbor is $\alpha$-removable; 
(3) non-removable: no neighbor is $\alpha$-removable, and at least two
neighbors are not $\beta$-removable. 

Let $\alpha$ and $\beta$ denote the probability that a random
neighbor of a random node $v$ in a network $\mathcal{G}$ is
$\alpha$-removable and $\beta$-removable in $\mathcal{G} \setminus v$,
respectively. We can derive two self-consistent
equations about $\alpha$ and $\beta$ 
\begin{eqnarray}
\label{eq:itilde-undir} \alpha &=& \sum_{k=1}^\infty Q(k) {\beta} ^{k-1} = A(1 - \beta),\\
\label{eq:l2tilde-undir} 1 - \beta &=& \sum_{k=1}^\infty Q(k) (1-\alpha) ^{k-1} = A(\alpha)
\end{eqnarray}
where $Q(k)\equiv k P(k)/c$ is the degree distribution for the node at
a random end of a randomly chosen edge, $c\equiv \sum_{k=0}^\infty k P(k)$
is the mean degree, and $A(x) \equiv \sum_{k=0}^\infty Q(k + 1) (1 -
x)^k$. %
These two equations indicate that
$\alpha$ satisfies  $x = A\big(A(x)\big)$. It can be shown that $\alpha$ is
the smallest fixpoint of $A\big(A(x)\big)$, i.e. the smallest root of the
function $f(x) \equiv A\big(A(x)\big) -x$. 

The expected fraction of non-removable nodes, i.e. the normalized core
size ($n_\mm{core}\equiv N_\mm{core}/N$), can then be calculated: 
\be
n_\mathrm{core} = \sum_{k=0}^\infty P(k) \sum_{s=2}^k \binom{k}{s}
\beta^{k-s} (1 - \beta - \alpha)^s, 
\label{eq:ncore}
\ee
which can be simplied in terms of
$G(x)\equiv \sum_{k=0}^\infty P(k) x^k$, i.e. the generating 
function of the degree distribution $P(k)$. The final result is given by 
\be
n_\mathrm{core} = G(1-\alpha) - G(\beta) - c\, (1-\beta
- \alpha) \, \alpha.
\label{eq:ncore-u}
\ee
For Erd\H{o}s-R\'enyi random networks, $G(x)=e^{-c(1-x)}=A(1-x)$, 
Eq.\ref{eq:ncore-u} can be further simplified as
$n_\mm{core}=(1-\beta-\alpha)(1-c\alpha)$, confirming previous
results\cite{Bauer-EPJB-01,Zhou-Book-12}.

The normalized number of edges in the core ($l_\mm{core}\equiv L_\mm{core}/N$) can also be calculated in
terms of $\alpha$ and $\beta$.  
Consider a uniform random edge, which remains in the core if and only if
both of its endpoints are non-removable without removing the edge. The
probability of one endpoint being non-removable without removing the
edge is $1 - \alpha - \beta$, and for the two endpoints the
probabilities are independent. Therefore, the expected normalized
number of edges in the core is 
\be
l_\mm{core} = \frac c2 \, ( 1-\alpha-\beta)^2. 
\label{eq:lcore-u}
\ee
with $c/2 = L/N$ the normalized number of edges in the network. Clearly, both $n_\mathrm{core} > 0$ and $l_\mm{core}>0$ if and only if $1- \beta - \alpha > 0$.

Now we consider directed networks $\mathcal{G}$ with given in- and
out-degree distributions, denoted by $P^-(k)$ and $P^+(k)$, respectively. 
Let $c$ denote the mean degree of each partition in the bipartite graph
representation $\mathcal{B}$ of the directed network $\mathcal{G}$, i.e. the
mean in-degree (or out-degree) of $\mathcal{G}$. 
Define $Q^\pm(k) \equiv k P^\pm(k) / c$, which is the degree distribution
of the upper or lower end, respectively, of a random edge in
$\mathcal{B}$. Define $A^\pm(x) \equiv \sum_{k=0}^\infty Q^\pm(k + 1) (1
- x)^k$. Then the same argument as we used in the undirected case
gives that 
\begin{eqnarray}
 \alpha^{\pm} &=& A^\pm (1 - \beta^\mp),\\
 1 - \beta^{\pm} &=& A^\pm (\alpha^\mp) 
\end{eqnarray}
and $\alpha^{\pm}$ is the
smallest fixpoint of %
$A^\pm \big(A^\mp(x) \big)$. Now we can
calculate the size of the core for each partition in $\mathcal{B}$ as  
\be
n_\mathrm{core}^\pm = \sum_{k=0}^\infty P^\pm(k) \sum_{s=2}^k \binom{k}{s}
(\beta^\mp)^{k-s} (1 - \beta^\mp - \alpha^\mp)^s
\label{eq:ncore-d1}
\ee
and we define the size of the core in the directed network $\mathcal{G}$ as 
\be
n_\mathrm{core} = (n_\mathrm{core}^+ + n_\mathrm{core}^-) / 2.
\label{eq:ncore-d2}
\ee
The normalized number of edges in the core can also be calculated
\be
l_\mm{core} = c \, ( 1- \alpha^+ - \beta^+) ( 1- \alpha^- - \beta^-). 
\label{eq:lcore-d}
\ee
\section{Condition for core percolation} 
\begin{figure}
\includegraphics[width=\textwidth]{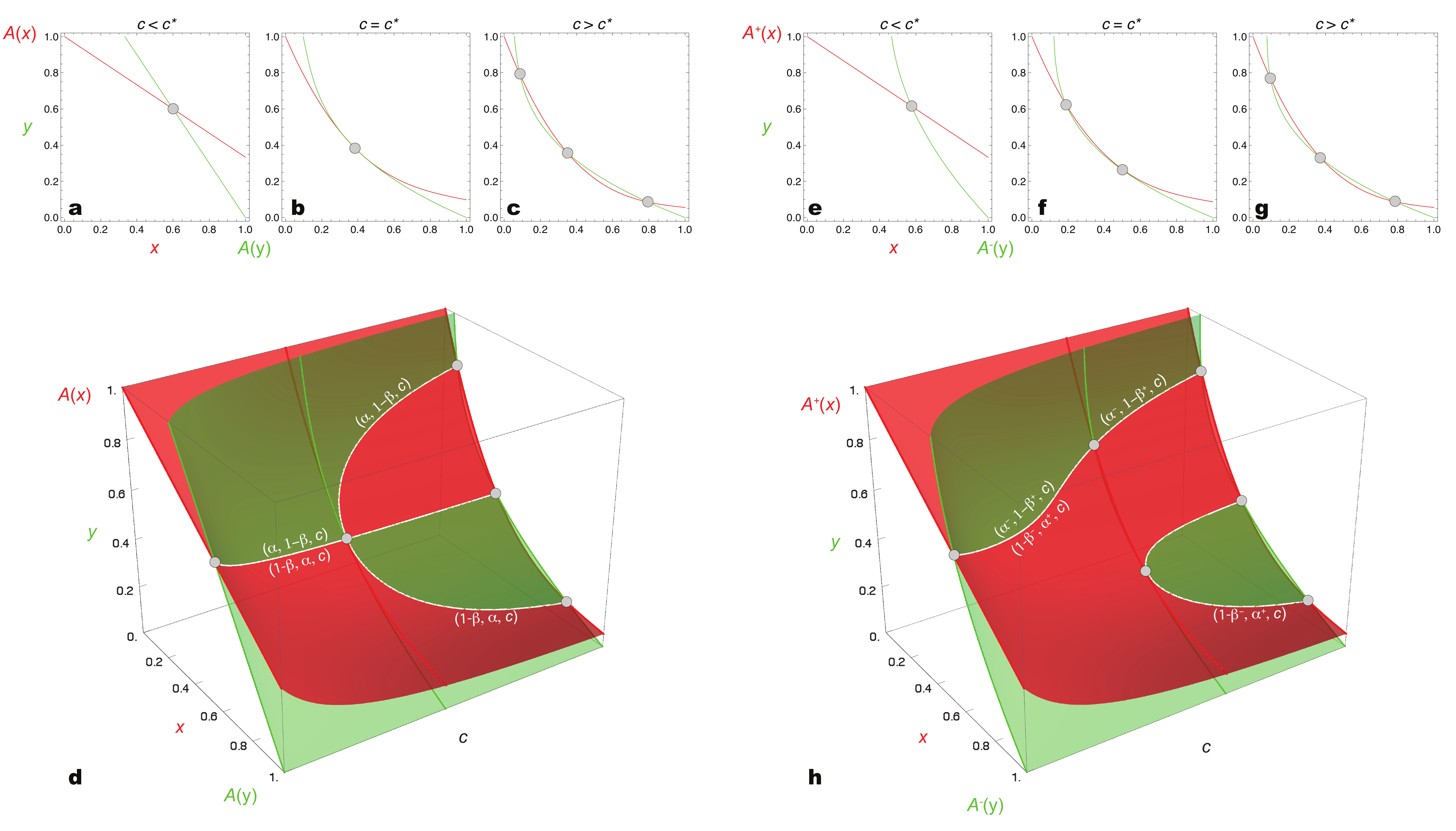}
\caption{{\bf Graphical solution of the self-consistent equations. } 
  {\bf a-d}, For undirected networks, the function $A(x)$ transforms
  the roots of $f(x)$ to the roots of the same function $f(x)$. 
  The graphical solution of $f(x)=A\big(A(x)\big)-x=0$ is best illustrated by
  plotting the two curves $A(x)$ vs. $x$ (in red) and $y$ vs. $A(y)$
  (in green) in the same coordinate system. 
  The coordinates of the intersection point(s) of the two curves give the
  solution(s) of $f(x)=0$. 
  In {\bf a}, {\bf b}, and {\bf c}, we show the graphical solutions
  for $c <, =,$ and $> c^*$, respectively. 
  {\bf d}, By drawing the two curves ($A(x)$ vs. $x$) and ($y$
  vs. $A(y)$) at different mean degrees $c$, we get two surfaces. The
  intersection curve of the two surfaces yields the solutions of
  $f(x)=0$ at different $c$ values. 
  For $c<c^*$, the intersection curve has one branch given by
  $(\alpha, 1-\beta, c) = (1-\beta, \alpha, c)$. 
  For $c>c^*$, the intersection curve has three branches. The top and
  bottom branches are given by $(\alpha, 1-\beta, c)$ and $(1-\beta,
  \alpha, c)$, respectively.   
  {\bf e-h}, For directed networks, $A^{\pm}(x)$ transforms the roots of
  $f^{\mp}(x)$ to the roots of $f^{\pm}(x)$. 
  The graphical solution of $f^{\pm}(x)=A^{\pm}(A^\mp(x))-x=0$ can be
  illustrated by plotting $A^+(x)$ vs. $x$ (in red) and $y$ vs. $A^-(y)$
  (in green) in the same coordinate system. 
  The $x$-coordinate (or $y$-coordinate) of the intersection point(s)
  of the two curves give the solution(s) of the 
  equation $f^-(x)=0$ (or $f^+(x)=0$,
  respectively). 
  In {\bf e}, {\bf f}, and {\bf g}, we show the graphical solutions
  for $c <, =,$ and $> c^*$, respectively. 
  {\bf h}, By drawing the two curves ($A^+(x)$ vs. $x$) and ($y$
  vs. $A^-(y)$) at different mean degrees $c$, we get two surfaces. The
  intersection curve of the two surfaces yields the solutions of
  $f^{\pm}(x)=0$ at different $c$ values. 
  For $c<c^*$, the intersection curve has one branch given by
  $(\alpha^-, 1-\beta^+, c) = (1-\beta^-, \alpha^+, c)$. 
  For $c>c^*$, the intersection curve has three branches. The top and
  bottom branches are given by $(\alpha^-, 1-\beta^+, c)$ and $(1-\beta^-,
  \alpha^+, c)$, respectively.   
}\label{fig:AAx}
\end{figure}

It is easy to see that the core in a
undirected network with degree distribution $P(k)$ is the very same as
in a directed network with the same out- and in-degree distributions,
i.e. $P^+(k) = P^-(k) = P (k)$. Therefore we can deal with directed
network for generality.  
As $n_\mm{core}$ is a continuous function of $\alpha^\pm$, we focus 
on $\alpha^{\pm}$, which is the smallest %
root of the function $f^{\pm}(x) \equiv A^{\pm}\big(A^{\mp}(x)\big) - x$.
There are several interesting facts about the function
$f^\pm(x)$. 
First of all, since $A^\pm(x)$ is a monotonically
decreasing function for $x\in [0,1]$ and $A^\pm(0)=1$ is the maximum
(see Figs.\ref{fig:AAx}, \ref{fig:analytic}), we have $f^\pm(0) > 0$ and $f^\pm(1) < 0$ (see
Fig.\ref{fig:analytic}c,d). Consequently, 
the number of 
roots (with multiplicity) of $f^\pm(x)$ in $[0, 1]$ is odd, and
numerical calculations suggest that this number is either 1 or 3
(see Figs.\ref{fig:AAx}, \ref{fig:analytic}). 
Secondly, if $f^\pm(x_0) = 0$ then $f^\mp\big(A^\mp(x_0)\big) = 0$, which
means $A^\mp(x)$ transforms the roots of $f^\pm(x)$ to the roots of
$f^\mp(x)$. 
This also suggests that $f^\pm(x)$ \emph{always} has a
trivial root $\alpha^\pm=A^\pm(\alpha^\mp)=1-\beta^\pm$. (For
undirected networks, $f(x)$ \emph{always} has a
trivial root $\alpha=A(\alpha)=1-\beta$.) 
Since $A^\mp(x)$ is a monotonically decreasing function and $\alpha^\pm$ is the smallest root of $f^\pm(x)$, $A^\mp(\alpha^\pm) = 1 -
\beta^\mp $ is therefore the largest root of $f^\mp(x)$.  
Hence $1 - \beta^\pm - \alpha^\pm$ is the
difference between the largest and the smallest roots of
$f^\pm(x)$ (see Fig.\ref{fig:AAx}). Consequently, if $f^\pm(x)$ has only one root
(which then must be the trivial root $\alpha^\pm=A^\pm(\alpha^\mp)=1-\beta^\pm$), then $1 -
\beta^\pm - \alpha^\pm=0$. According to Eq.\ref{eq:ncore-d1}, this implies that there is no core. On the other hand, if
multiple roots exist and they are different then $1 - \beta^\pm -
\alpha^\pm > 0$, and the
core will develop. %

We apply the above condition to the following random undirected
networks with specific degree distributions\cite{Newman-PRE-01}.  
(1) Erd\H{o}s-R\'enyi (ER)\cite{Erdos-PMIHAS-60,Bollobas-Book-01}
networks with Poisson degree distribution $P(k)=e^{-c} c^k /
k!$, $A(x)=e^{-cx}$ and $f(x)=\exp(-c e^{-cx})-x$. As shown in
Fig.\ref{fig:analytic}a, the core percolation occurs at $c=c^*=e$,
which agrees with previous theoretical
results\cite{Karp-IEEE-81,Bauer-EPJB-01,Zhou-Book-12}. 
(2) Exponentially distributed graphs with $P(k)=(1-e^{-1/\kappa})
e^{-k/\kappa}$ and mean degree $c=e^{-1/\kappa}/(1-e^{-1/\kappa})$. We
find that core percolation occurs at $c=c^*=4$. 
(3) Purely power-law distributed networks with
$P(k)=k^{-\gamma}/\zeta(\gamma)$ for $k\ge 1$, $\gamma>2$ and $\zeta(\gamma)$ the Riemann $\zeta$ function. We find
that $f(x)$ has no multiple roots and hence $n_\mm{core}=0$ for all
$\gamma>2$. In other words, for purely  scale-free (SF) networks, the core
does not exist. %
(4) Power-law distributed networks with exponential degree cutoff, i.e. $P(k)=\frac{
  k^{-\gamma} \, e^{-k/\kappa}}{\mm{Li}_{\gamma}(e^{-1/\kappa})}$ for
$k\ge 1$ with $\mm{Li}_{n}(x)$ the $n$th polylogarithm of $x$. We find
that $n_\mm{core}=0$ for $\gamma>\gamma_\mm{c}(\kappa)$, and the threshold
value $\gamma_\mm{c}(\kappa)$ approaches 1 as $\kappa$ increases. 
Hence, for SF networks with exponential
degree cutoff the core still does not exist for all $\gamma>1$. %
(5) Asymptotically SF networks generated by the static model with  
$ P(k) = \frac{[\frac c2 (1-\xi)]^{1/ \xi}}{\xi}
\frac{\Gamma(k-1/ \xi,\frac c2 [1-\xi])}{\Gamma(k+1)}$, where $\Gamma(s)$ is the
gamma function and $\Gamma(s,x)$ the upper incomplete gamma
function\cite{Goh-PRL-01,Catanzaro-EPJB-05,Lee-EPJB-06}. 
In the large $k$ limit, $P(k) \sim k^{-(1+\frac{1}{\xi})} =
k^{-\gamma}$ where $\gamma= 1 + \frac{1}{\xi} > 2$. 
For small $k$, $P(k)$ deviates significantly from the power-law
distribution\cite{Catanzaro-EPJB-05} and there are much fewer small-degree nodes than the purely scale-free networks, which results in a
drastically different core percolation behavior. 

Hereafter, we systematically study the net effect of adding more links
(i.e. increasing mean degree $c$, yet without changing other
parameters in $P(k)$) on core percolation. ER networks and the
asymptotically SF networks generated by the static model naturally serve
this purpose, since their mean-degree is an independent and explicit
tuning parameter. 

\section{Nature of core percolation} 
\begin{figure}
\includegraphics[width=\textwidth]{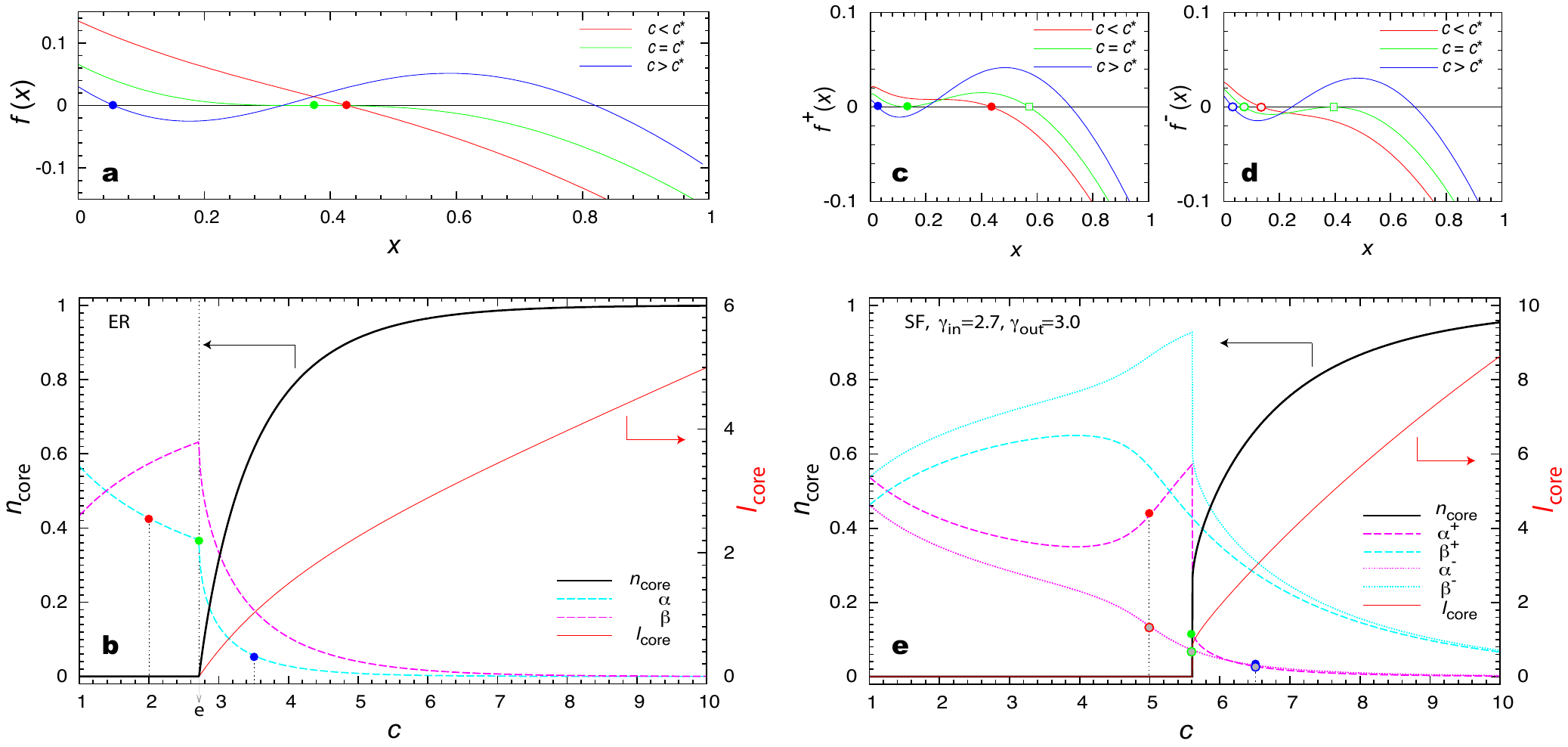}
\caption{{\bf Analytical solution of the core percolation.} 
  {\bf a-b}, Undirected Erd\H{o}s-R\'enyi (ER) random networks. 
  {\bf a}, $\alpha$ is the smallest root of the function $f(x)\equiv A\big(A(x)
  \big) -x$, represented by red, green, and blue dots for $c <, =,$
  and $> c^* = e$, respectively. 
  {\bf b}, $\alpha, \beta, n_\mm{core}$ and $l_\mm{core}$ as functions
  of the mean degree $c$.
  {\bf c-e}, Directed asymptotically scale-free (SF) random networks
  generated by the static model. Both the in-degree and
  out-degree distributions of the networks are scale-free with
  degree exponents $\gamma_\mm{in}=2.7$ and $\gamma_\mm{out}=3.0$.
  {\bf c, d}, $\alpha^{\pm}$ is the smallest root of the function
  $f^{\pm}(x)\equiv A^{\pm}\big(A^{\mp}(x)\big) - x$, represented by red,
  green, and blue dots for $c <, =,$ and $> c^* \simeq 11.2$,
  respectively. 
  {\bf e}, $\alpha^\pm, \beta^\pm, n_\mm{core}$ and $l_\mm{core}$ as functions
  of the mean degree $c$. 
  The jumps in $\alpha^+$ and $\beta^-$ result in the jumps in
  $n_\mm{core}$ and $l_\mm{core}$, hence the first-order core percolation
  occurs. 
\vspace{5in}
}\label{fig:analytic}
\end{figure}

We observed that if the mean degree $c$ is small, then
$f^\pm(x)$ has one root, but if $c$ is large, $f^\pm(x)$ has three
roots (see Figs.\ref{fig:AAx}, \ref{fig:analytic}). At the critical point $c=c^*$, the
number of roots jumps from 1 to 3 by the appearance of one new root with
multiplicity 2. (Note that $f^\pm(x)$ cannot immediately intersect the
$x$-axis at two new points, but it touches first.) 
This explains why the core percolation occurs at $c=c^*$.  

According to the transformation from the roots of $f^\pm(x)$ to the
roots of $f^\mp(x)$ through $A^\mp(x)$, 
for either $f^+(x)$ or $f^-(x)$ (depending on the details of $P^+(k)$
and $P^-(k)$) its new root at
$c=c^*$ is smaller than its original root; and
for either $f^-(x)$ or $f^+(x)$ the new root at
$c=c^*$ is larger than the original root; 
or there is a degenerate case when this new root is the same as the
original root for both $f^+(x)$ and $f^-(x)$. 
For example, for directed asymptotically SF networks generated by the static model with $\gamma_\mm{in}=2.7,
\gamma_\mm{out}=3.0$, the new root (marked as green dot) of
$f^+(x)$ at $c=c^*$ is smaller than the original root (green
square) of $f^+(x)$  (see Fig.\ref{fig:analytic}c), and the new root (green square) of
$f^-(x)$ at $c=c^*$ is larger than the original root (green circle) of
$f^-(x)$ (see Fig.\ref{fig:analytic}d). 
In other words, at the critical point, for either $f^+(x)$ or $f^-(x)$, its smallest two
roots are the same, and for the other function (either $f^-(x)$ or
$f^+(x)$), its largest two roots are the same (see Fig.\ref{fig:analytic}c,d).  
While for directed networks with $P^+(k)=P^-(k)=P(k)$, i.e. the
degenerate case, we have 
$f^+(x)=f^-(x)=f(x)$, and the new root of $f(x)$ at $c=c^*$ has to be the
same as the the original root of $f(x)$, i.e. all three
roots must be the same (see Fig.\ref{fig:analytic}a).
Therefore at the critical point, unless in the degenerate case,
$\alpha^+$ together with $\beta^-$ (or $\alpha^-$ together with
$\beta^+$) decrease discontinuously, which implies a discontinuous
transition in the core size. 
To sum up, in the degenerate case that
$P^+(k)=P^-(k)=P(k)$ core percolation is continuous, but for general
non-degenerate case $P^+(k) \neq P^-(k)$, we have a discontinuous 
transition in both $n_\mm{core}$ and $l_\mm{core}$. These results are
clearly shown in Fig.\ref{fig:analytic}b,e.

\begin{figure}
\begin{center}
\includegraphics[width=0.75\textwidth]{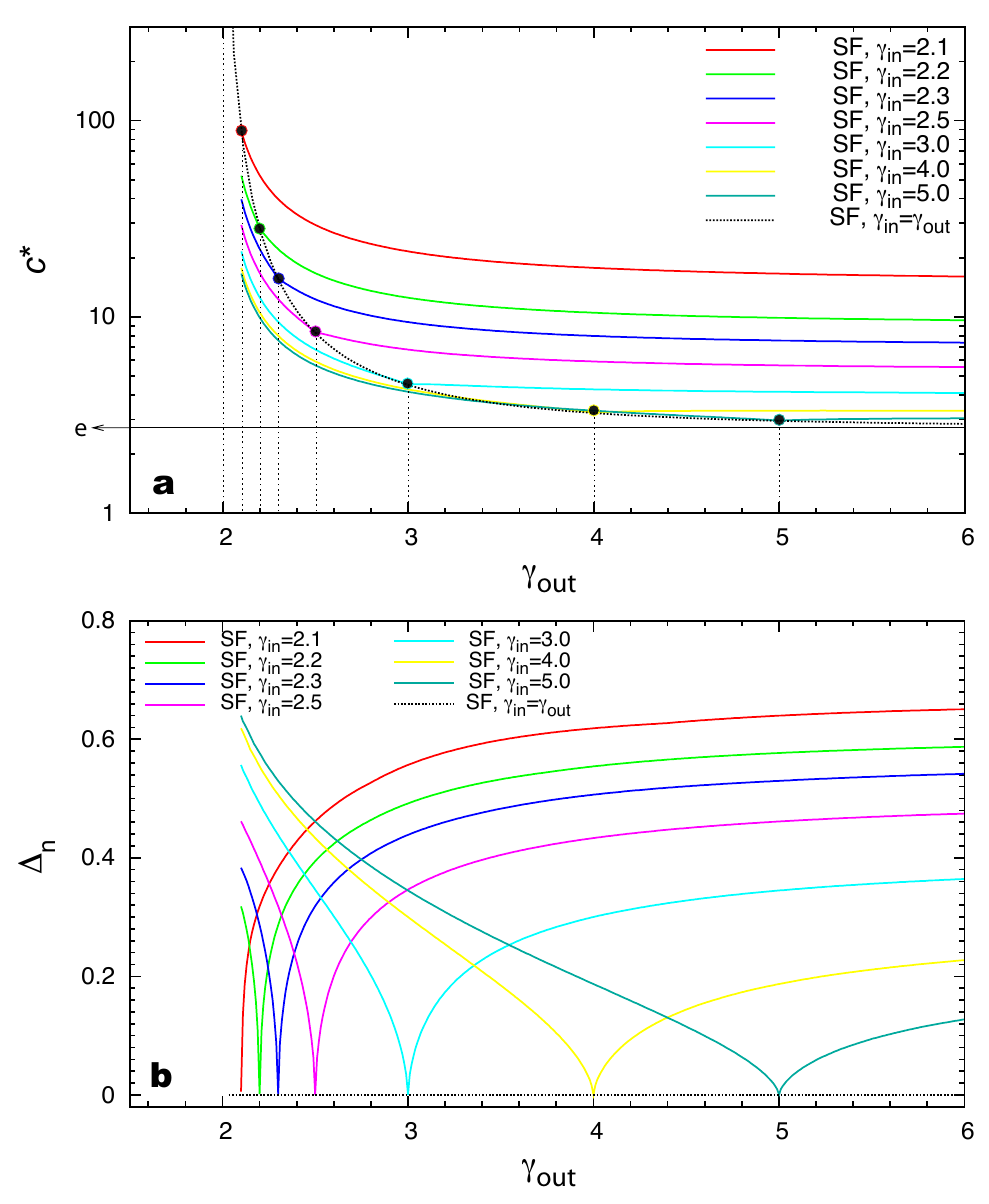}
\end{center}
\caption{{\bf Threshold and discontinuity of core percolation.} 
  {\bf a}, Analytical solution of the core percolation threshold $c^*$
  calculated by solving $f^\pm(x)={f^\pm}'(x)=0$ for model networks. For
  ER networks, $c^*=e$. For undirected asymptotically SF networks generated by the
  static model, $c^* \to \infty$
  as $\gamma \to 2$, and and $c^* \to e$ as $\gamma \to \infty$. 
  {\bf b}, The discontinuity $\Delta_\mm{n}$ in $n_\mm{core}$ at $c=c^*$ for
  model networks. For undirected or directed networks with
  $P^+(k)=P^-(k)$, $\Delta_\mm{n}=0$. For directed network, $\Delta_\mm{n}$
  increases as the difference between the in- and out-degree
  distributions (quantified by the difference between the degree
  exponents $\gamma_\mm{in}$ and $\gamma_\mm{out}$) increases. 
\vspace{3in}
}\label{fig:kc_Delta}
\end{figure}

At the critical point $c^*$, $f^\pm(x)$ touches the $x$-axis at its
new root (see Fig.\ref{fig:analytic}c,d), hence we have either $f^+(\alpha^+) =
(f^+)'(\alpha^+) = 0$ (or $f^-(1-\beta^-) = (f^-)'(1-\beta^-) = 0$),
which enable us to calculate the core percolation threshold $c^*$. 
In the degenerate case, if $c\le c^*$ then
$f(\alpha)= f'(\alpha) = 0$ can be further simplified as
$A(\alpha)=\alpha$ and $[A'(\alpha)]^2=1$.    
The results of $c^*$ for ER and SF networks generated by the static
model are shown in Fig.\ref{fig:kc_Delta}a.

The discontinuity in $n_\mm{core}$ and
$l_\mm{core}$ at $c^*$, denoted by $\Delta_\mm{n}$ and
$\Delta_\mm{l}$ respectively, can also be calculated 
\bea
\Delta_\mm{n} &=& \frac 12 \left(\Delta_\mm{n}^+ + \Delta_\mm{n}^-
\right)\\
\Delta_\mm{l} &=& c^*  (1-\beta^{-,*}
- \alpha^{-,*})  (1-\beta^{+,*}
- \alpha^{+,*}) 
\eea
with $\Delta_\mm{n}^\pm\equiv G^\pm(1-\alpha^{\mp,*}) -
G^\pm(\beta^{\mp,*}) - c^*\,  (1-\beta^{\mp,*}
- \alpha^{\mp,*}) 
\, \alpha^{\pm,*}$. %
The results of $\Delta_\mm{n}$ for ER and SF networks generated by the static
model are shown in Fig.\ref{fig:kc_Delta}b. 
We find that
$\Delta_\mm{n} \to 0$ as $\gamma_\mm{in} \to \gamma_\mm{out}$, consistent
with the result obtained above that core percolation is
continuous for undirected networks or directed networks with
$P^+(k)=P^-(k)$.  
We also find that $\Delta_\mm{n}$ increases as the differences between
$\gamma_\mm{in}$ and $\gamma_\mm{out}$ increases. 

We can further show that in the general non-degenerate case, core
percolation is actually a hybrid phase transition\cite{Parisi-PRE-08,
  Dorogovtsev-PRL-06, Goltsev-PRE-06}, i.e. $n_\mm{core}$ (or
$l_\mm{core}$) has a jump at the critical point as at a first-order
phase transition but also has a critical singularity as at a
continuous transition. The results are summarized here: %
in the critical regime $\epsilon  = c-c^* \to 0^+$ 
\bea
n_\mm{core} - \Delta_\mm{n} &\sim& (c-c^*)^\eta \\
l_\mm{core}  - \Delta_\mm{l} &\sim& (c-c^*)^\theta 
\eea
with the critical exponents $\eta = \theta = \frac 12$. 
Our calculations do not use any specific functional form of
$A^\pm(x)$. Instead, we only assume that they are continuous functions
of the mean degree $c$. 
Interestingly, in the degenerate or undirected case, one has a continuous phase
transition ($\Delta_\mm{n}=\Delta_\mm{l}=0$) but with a completely different
set of critical exponents: $\eta' = \theta' = 1$ \cite{Bauer-EPJB-01}. 
\section{Numerical results}
\begin{figure}
\includegraphics[width=\textwidth]{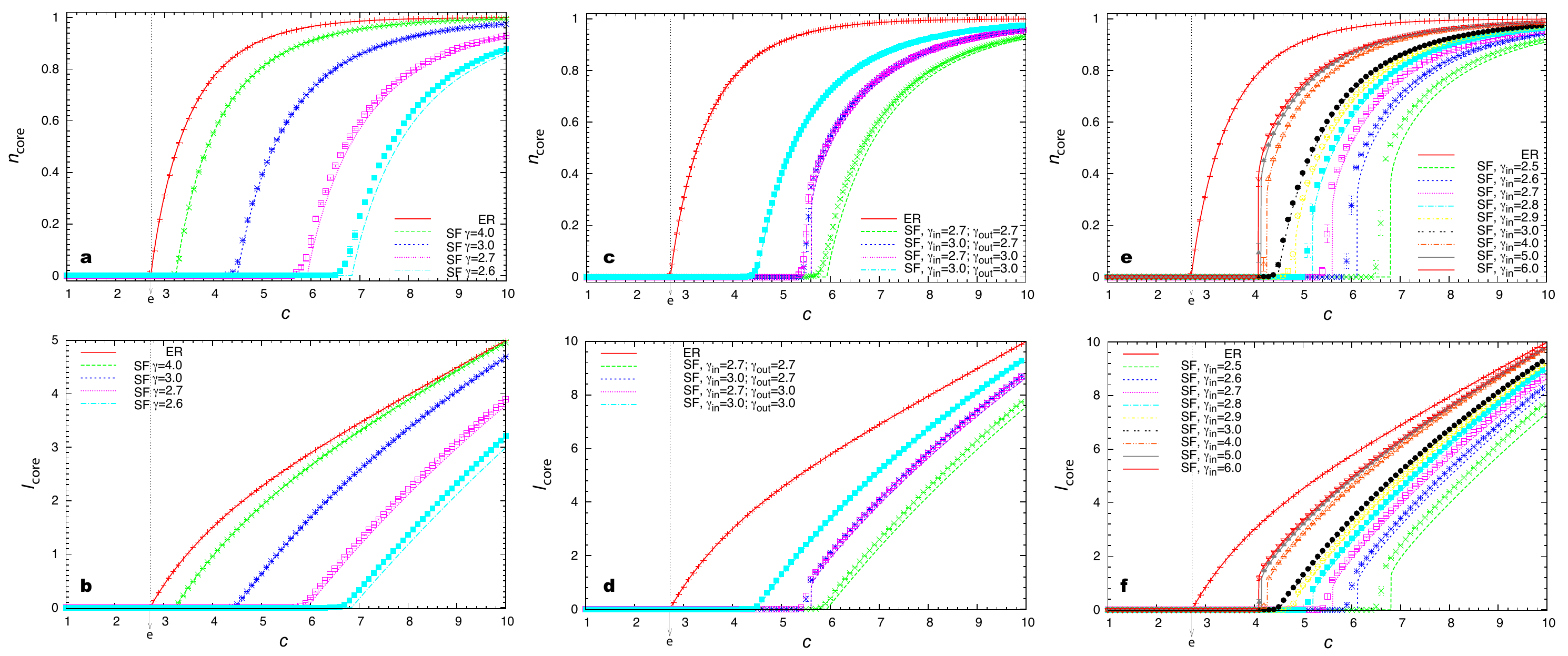}
\caption{{\bf Core percolation in random networks.} 
  Symbols are numerical results calculated from the GLR
  procedure on finite discrete networks constructed with the static
  model\cite{Goh-PRL-01} with $N=10^5$. The numerical results are
  averaged over 20 realizations with error bars defined as s.e.m. 
  Lines are analytical results for infinite large system ($N\to
  \infty$) calculated from Eq.\ref{eq:ncore-u} and \ref{eq:lcore-u}
  for undirected networks or Eq.\ref{eq:ncore-d2} and \ref{eq:lcore-d}
  for directed networks.  
  Finite size effects are more discernable for $\gamma\to 2$, which can be
  eliminated by imposing degree cutoff in constructing the SF
  networks\cite{Boguna-EPJB-04, Chung-AC-02}. 
  {\bf a-b}, The normalized core size ($n_\mm{core}=N_\mm{core}/N$)
  and the normalized number of edges in the core
  ($l_\mm{core}=L_\mm{core}/N$) for undirected model networks:
  Erd\H{o}s-R\'enyi (ER) and asymptotically scale-free (SF) with different values of
  $\gamma$.  
  For both model networks, the core percolation is
  continuous, which is fundamentally different from the $k\ge 3$-core
  percolation, which becomes discontinuous for ER networks and SF
  networks with $\gamma > 3$\cite{Pittel-JCT-96,Dorogovtsev-PRL-06}.  
  {\bf c-d}, $n_\mm{core}$  and $l_\mm{core}$ for directed ER and
  asymptotically SF model networks. 
  The core percolation is continuous if the out- and in-degree
  distributions are the same ($P^+(k)=P^=(k)$) while it
  becomes discontinuous if $P^+(k)\neq P^=(k)$. 
  {\bf c-d}, For directed SF networks with fixed $\gamma_\mm{out} =
  3.0$, by tuning $\gamma_\mm{in}$ we see that the discontinuity in
  both $n_\mm{core}$ and $l_\mm{core}$ become larger as the difference
  between $\gamma_\mm{in}$ and $\gamma_\mm{out}$ increases. 
\vspace{3in}
}\label{fig:numeric}
\end{figure}

We check our analytical results with extensive numerical calculations
by performing the GLR procedure on finite discrete networks generated
by the static model\cite{Goh-PRL-01,Catanzaro-EPJB-05,Lee-EPJB-06}.  
Fig.\ref{fig:numeric}a and \ref{fig:numeric}b show $n_\mm{core}$ and $l_\mm{core}$ (in symbols) for
undirected ER networks and asymptotically SF networks with different
degree exponents. For comparison, analytical results for infinite
large networks are also shown (in lines). 
Clearly, core percolation is continuous in this case.    
This is fundamentally different from the $k\ge 3$-core
percolation, which becomes discontinuous for ER networks and SF
networks with $\gamma > 3$\cite{Pittel-JCT-96,Dorogovtsev-PRL-06}.  

Fig.\ref{fig:numeric}c and \ref{fig:numeric}d show the results of $n_\mm{core}$ and $l_\mm{core}$
for directed networks. 
For directed networks with the same in- and out-degree
distributions, e.g. directed ER networks or directed SF networks
with $\gamma_\mm{in} = \gamma_\mm{out}$ generated by the static
model, the core percolation is still continuous. 
But for directed networks with different in- and out-degree
distributions, e.g. directed SF networks with $\gamma_\mm{in} \neq
\gamma_\mm{out}$ generated by the static
model, the core percolation looks discontinuous. The
discontinuity in $n_\mm{core}$ (or $l_\mm{core}$) increases as
the difference  between $\gamma_\mm{in}$ and $\gamma_\mm{out}$
increases (see Fig.\ref{fig:numeric}e,f).

\section{Real networks}
\begin{figure}[t!]
\begin{center}
\includegraphics[width=\textwidth]{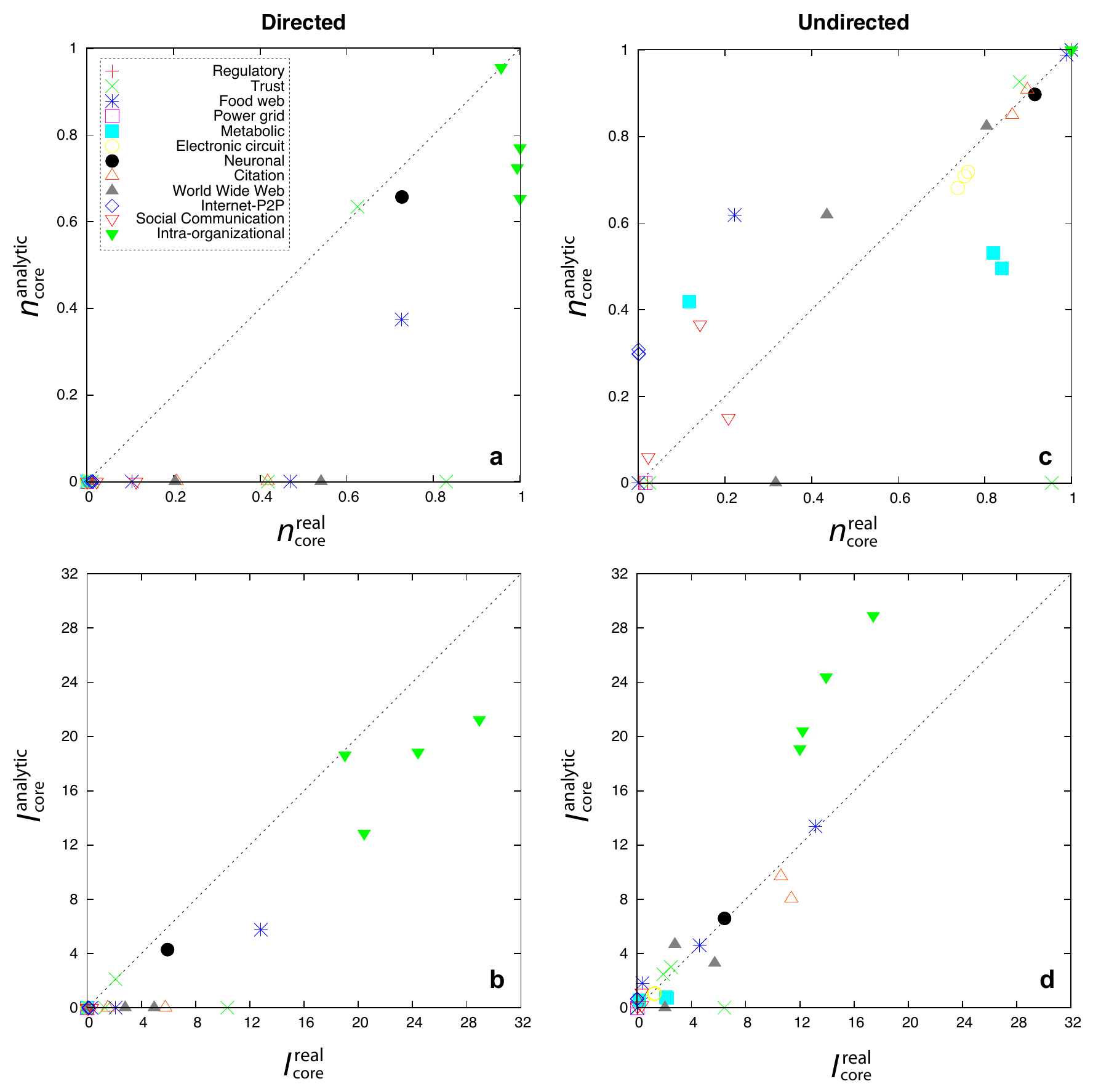}
\end{center}
\caption{{\bf Normalized core size for real networks, compared with
    analytical predictions.} All the real networks
  considered here are directed. For data sources and references, see Ref.\cite{Liu-Nature-11} Supplementary Information Sec.VI.  {\bf a-b}, By applying the
  GLR procedure we yield 
  $n_\mm{core}^\mm{real}$ and $l_\mm{core}^\mm{real}$. Using
  Eq.\ref{eq:ncore-d2} and Eq.\ref{eq:lcore-d} with out- and in-degree
  distributions ($P^+(k)$ and $P^-(k)$) as the only inputs, we obtain
  $n_\mm{core}^\mm{analytic}$ and $l_\mm{core}^\mm{analytic}$. 
  {\bf c-d} By ignoring the direction of the edges, we can treat the
  original directed networks as undirected ones and apply the GLR
  procedure to get $n_\mm{core}^\mm{real}$ and
  $l_\mm{core}^\mm{real}$. Similarly, we can calculate
  $n_\mm{core}^\mm{analytic}$ and $l_\mm{core}^\mm{analytic}$ by using
  Eq.\ref{eq:ncore-u} and Eq.\ref{eq:lcore-u} with the degree
  distribution $P(k)$ as the only input.
\vspace{1in}
} \label{fig:real}
\end{figure}

We also apply our theory to real-world networks with known degree
distributions. In Fig.\ref{fig:real} we demonstrate that in some cases 
our analytical results calculated from Eqs.\ref{eq:ncore-u},
\ref{eq:lcore-u} (or Eqs.\ref{eq:ncore-d2}, \ref{eq:lcore-d}) with
degree distribution as the only input predict with surprising accuracy
the core size of real networks. Yet, in other cases there is a
noticeable difference between theory and reality, which suggests the
presence of extra structure in the real-world networks that is not
captured by the degree distribution. 
In particular we find that almost all the directed real-world networks have
larger core sizes than the theoretical predictions (see
Fig.\ref{fig:real}a,b). In other words, those networks are
``overcored''. While if we treat those networks as undirected ones,
their core sizes deviate from our theory in a more complicated manner. 
The effects of higher order correlations (e.g. degree
correlations\cite{Newman-PRL-02}, clustering\cite{Watts-Nature-98},
loop structure\cite{Bianconi-PRL-08} and
modularity\cite{Newman-PNAS-06}) may play very important roles to
explain the discrepancy between theory and reality.

\section{Conclusion} 
In sum, we analytically solve the core percolation problem in both
undirected and directed random networks with arbitrary degree
distributions. We show the condition for core percolation. We find it
is continuous in undirected networks (if it occurs), while it becomes
discontinuous or hybrid in directed networks unless the in- and
out-degree distributions are the same.  
Within each case, the critical exponents associated with the critical
singularity are universal for random networks with arbitrary degree
distributions parameterized continuously in mean degree. 
But the two cases have totally different sets of critical
exponents. 
These results vividly illustrate that core percolation is a
fundamental structural transition in complex networks and its
implication on other problems, e.g. conductor-insulator transitions, 
combinatorial optimization problems, and network controllability
issue, deserves further exploration. 
The analytical framework presented here also raises a number of questions,
answers to which would further improve our understanding of core
percolation on complex real-world networks. For example, we focused on
uncorrelated random networks and leave the systematic studies of the
effects of higher order correlations as future work. 

\vspace{0.2in}
{\small 
We thank Braden Brinkman and Chaoming Song for valuable
discussions. This work was partially supported by the Network
Science Collaborative Technology Alliance sponsored by the US Army
Research Laboratory under Agreement Number W911NF-09-2-0053; the
Office of Naval Research under Agreement Number N000141010968; the
Defense Threat Reduction Agency awards WMD BRBAA07-J-2-0035 and
BRBAA08-Per4-C-2-0033; the James S. McDonnell Foundation 21st
Century Initiative in Studying Complex Systems; MTA R\'enyi ``Lendulet''
Groups and Graphs Research Group; ERC Advanced Research Grant
No. 227701 and KTIA-OTKA grant No. 77780. 
All authors have contributed equally to this
work. Correspondence and requests for materials should be addressed to
Yang-Yu Liu ~(email: ya.liu@neu.edu). 
}

\newpage

\newpage

\newpage

\newpage

\newpage

\newpage

\newpage

\end{document}